\documentstyle[sprocl,epsfig]{article}

\def\Journal#1#2#3#4{{#1} {\bf #2}, #3 (#4)}

\def\NPA{{\em Nucl. Phys.} A}
\def\PLB{{\em Phys. Lett.}  B}
\def\PRL{\em Phys. Rev. Lett.}

\def\PRC{{\em Phys. Rev.} C}

\def\ANN{\em Ann. Phys. (N.Y.)}

\def\be{\begin{equation}}
\def\ee{\end{equation}}
\def\bea{\begin{eqnarray}}
\def\eea{\end{eqnarray}}

\begin{document}

\title{WHAT CAN WE LEARN FROM NUCLEAR MATTER INSTABILITIES?}

\author{V. BARAN\footnote{on leave from NIPNE-HH, Bucharest, Romania}
, M. COLONNA, M. DI TORO}

\address{Laboratorio Nazionale del Sud, Via S. Sofia 44,
I-95123 Catania, Italy and NIPNE-HH, Bucharest, Romania}

\author{M. ZIELINSKA-PFABE}

\address{Smith College, Northampton, USA}

\author{H.H. WOLTER}

\address{Sektion Physik, University of Munich, Germany}


\maketitle\abstracts{ 
We discuss the features of instabilities in binary systems,
in particular, for asymmetric nuclear matter. We show its relevance for the
interpretation of results obtained in experiments and in "ab initio"
simulations of the reaction between $^{124}Sn+^{124}Sn$ at 50AMeV.}

\section{Instabilities in asymmetric nuclear matter}

The process of multifragmentation following the collision of heavy nuclei
in the region of medium energies displays several features analogous to
usual liquid-gas phase transitions of water. However in this
analogy one should by aware of differences due to Coulomb, finite size and
quantum effects as well as to the binary, i.e. two-component, character of 
nuclear matter. Moreover,
the time scales of the process are of the order of, or shorter than, the 
relaxation times of
the relevant degrees of freedom. Therefore we have to consider not only
equilibrium phase-transition in binary systems, but in an equally important way
 the dynamical evolution driving such phase transition and its
dependence on the symmetry term of the equation-of-state. Indeed after a fast
compression and expansion we expect the nuclear system to be quenched 
into an unstable state either inside the coexistence curve in the
metastability region (where the phase is unstable against short wave length but
large amplitude fluctuations) or in the instability (spinodal) region
(were the system becomes unstable against long wave length but small amplitude
fluctuations). 
  Then the system will evolve toward a stable thermodynamical state
 of two coexisting phases either through nucleation
(in former case) or through spinodal decomposition (in the latter case). These 
aspects were discussed repeatedly in the past,
\cite{Peth87,Peth88,Peth93,col94,cdg94,col98}
but the binary character will induce new features for both scenarios that are
absent in one-component systems.
Therefore studying the nature of instabilities which develop in nuclear
systems, their relaxation times and the influence
of the isospin degree of freedom will provide information on the mechanisms 
involved in the fragment formation processes and will set limits
for the applicability of fully equilibrium approaches.
  
     In the framework of Landau theory for two component Fermi liquids the
spinodal border was determined by studying the stability of collective modes 
described by two coupled Landau-Vlasov equations for protons and neutrons.
In terms of the appropriate Landau parameters the stability condition can be 
expressed as~\cite{bar98},
\begin{equation}
(1 + F_0^{nn})(1 + F_0^{pp}) - F_0^{np}F_0^{pn} > 0~.  
\label{eq:Land}
\end{equation}  
It is possible to show that this condition is equivalent to 
the following thermodynamical condition
\begin{equation}
\left({\partial P \over \partial \rho}\right)_{T,y}
\left({\partial\mu_p \over \partial y}\right)_{T,P} > 0~.
\label{eq:ser}                                                   
\end{equation}
discussed in \cite{landau,serot}, where $y$ is the proton fraction.       
In Fig. 1 we show the spinodal lines obtained from eq.~(\ref{eq:Land})
(continuous line with dots) which for asymmetric nuclear matter is seen to 
contain the lines corresponding to
"mechanical instability",
$\left({\partial P \over \partial \rho}\right)_{T,y}<0$
(crosses). Therefore eqs. (\ref{eq:Land},\ref{eq:ser}) describe the "chemical instability" of nuclear matter.
\begin{figure}[htb]
\epsfysize=4.5cm
\centerline{\epsfbox{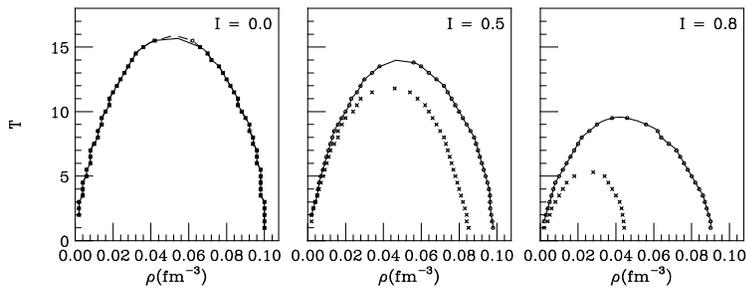}}
\caption
{
Spinodal lines corresponding to chemical (joined points) and
mechanical (crosses) instability for three asymmetries, I=(N-Z)/(N+Z)=0.0,0.5,0.8 
of nuclear matter.
}
\end{figure}

We want to stress, however, that by just looking at the above stability conditions
we cannot determine the nature of the fluctuations against which a binary
system becomes chemically unstable. Indeed, the thermodynamical
condition in eq.~(\ref{eq:ser}) cannot distinguish between two  very different 
situations which can be encountered in nature: an attractive
interaction between the two components of the mixture ($F_0^{np},F_0^{pn} < 0$),
 as is the case of nuclear matter, or a  repulsive interaction
between the two species.
We define as isoscalar-like density fluctuations
the case when proton and neutron Fermi spheres
(or equivalently the proton and neutron densities) fluctuate in phase and as
isovector-like density fluctuations when the two Fermi sphere fluctuate
out of phase. Then it is possible to prove, based on 
a thermodynamical approach of asymmetric Fermi liquid mixtures\cite{bar20},
that
chemical instabilities are triggered by isoscalar fluctuations in the first, 
i.e. attractive, situation
and by isovector fluctuations in the second one. For the asymmetric nuclear
matter case because of the attractive interaction between protons and neutrons 
the phase transition is thus due to isoscalar fluctuations that induce
chemical instabilities while the system is never unstable against isovector
fluctuations. 
Of course the same attractive interaction is also at the origin of phase
transitions in symmetric nuclear matter. However, in the asymmetric case
isoscalar fluctuations lead to a more symmetric high density phase
everywhere under the instability line defined by eq.(\ref{eq:Land})~\cite{bar98}.

\begin{figure}[htb]
\epsfysize=5.6cm
\centerline{\epsfbox{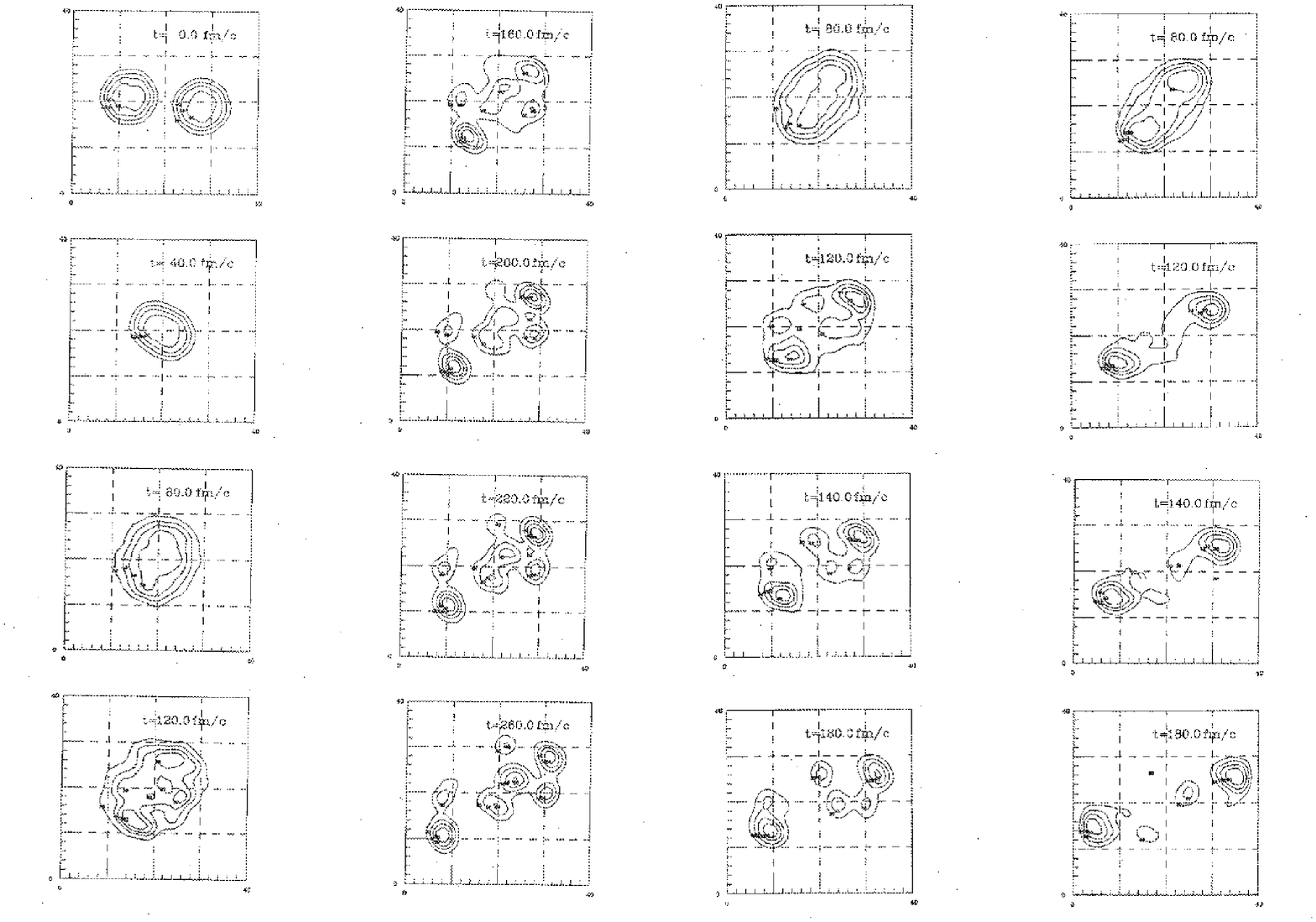}}
\caption{
$^{124}Sn+^{124}Sn$ $50AMeV$: time
evolution of the nucleon density projected on the reaction plane.
First two columns: $b=2fm$ collision, approaching, compression and
separation phases. Third and fourth columns: $b=4fm$ and $b=6fm$,
separation phase up to the {\it freeze-out}.
}
\end{figure}

\section{"Ab initio" simulations of Sn + Sn reactions}

A new code for the solution of microscopic transport equations,
the {\it Stochastic Iso-BNV},
has been written where asymmetry effects are
suitably accounted for and the dynamics of fluctuations is
 included \cite{fab98,flu98}.

We have studied the $50AMeV$ collisions of the systems
$^{124}Sn+^{124}Sn$ and $^{112}Sn+^{112}Sn$, where new data
have been measured at $NSCL-MSU$ \cite{betty}.
We will discuss averages over $100$ events generated in semi-central ($b=2fm$)
and semi-peripheral ($b=6fm$) reactions. In Fig.2 we show the  evolution 
with impact parameter of the density plot
(projected on the reaction plane) for one event each
(neutron rich case $^{124}Sn+^{124}Sn$, EOS with stiff asymmetry term) \cite{cris2}.

We remark: i) In the cluster formation we see a quite clear transition 
with impact parameter from
{\it bulk}  \cite{col94,cdg94,col98} to {\it neck} \cite{col95,inri97,dur98}
instabilities. ii) The {\it "freeze-out times"}, i.e. when the nuclear
interaction among clusters disappears, are decreasing with impact parameter.
These two dynamical
effects will influence the isospin content of the produced primary
fragments, as shown below.

\begin{figure}[htb]
\epsfysize=4.6cm
\centerline{\epsfbox{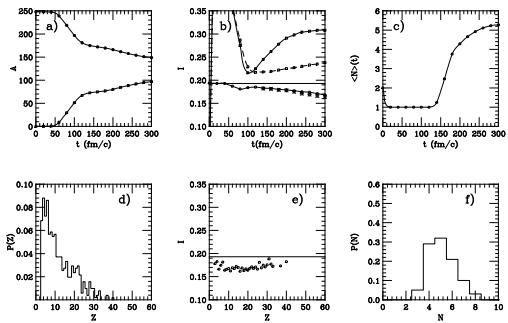}}
\caption
{$^{124}Sn+^{124}Sn$ $50AMeV$ $b=2fm$ collisions: time
evolution and freeze-out properties.
See text.
}
\end{figure}

\begin{figure}[htb]
\epsfysize=4.6cm
\centerline{\epsfbox{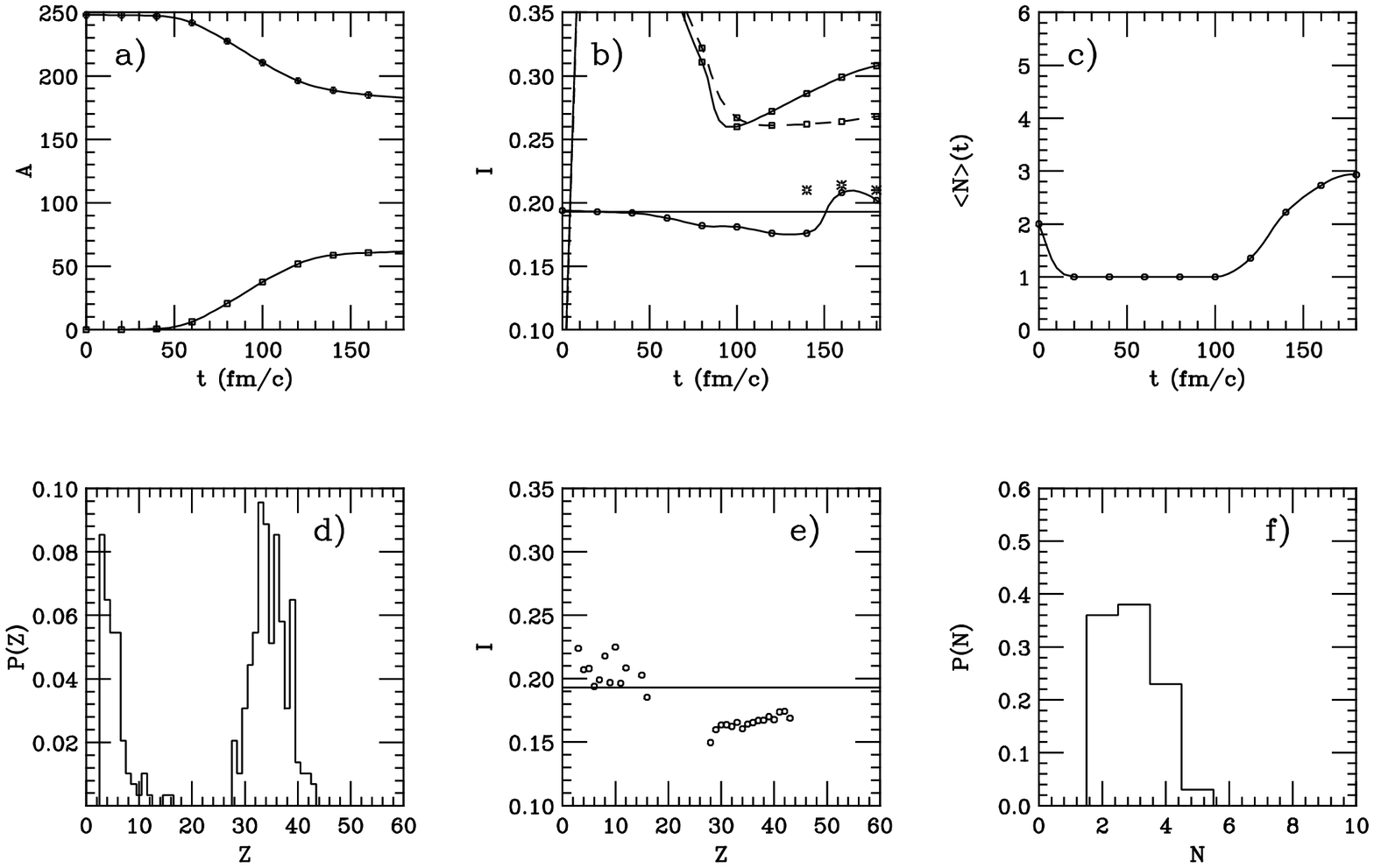}}
\caption
{$^{124}Sn+^{124}Sn$ $50AMeV$ $b=6fm$ collisions: time
evolution and freeze-out properties.
See text.
}
\end{figure}

A detailed analysis of the results for the same system
 is shown in Fig.3 (central, $b=2fm$)
and Fig.4 (peripheral, $b=6fm$). Each figure is organized in the following way:
Top row, time evolution of: (a) {\it Mass} in the
liquid (clusters with  $Z \geq 3$, upper curve) and the gas (lower curve) phase; 
(b) {\it Asymmetry} $I=(N-Z)/(N+Z)$ in gas "central" (solid line with squares),
gas total (dashed+squares), liquid
"central" (solid+circles) and IMF's ($3 \leq Z \leq 12$, stars).
"Central" means in a cubic box of side $20fm$ around the c.m..
The horizontal line shows the initial
average asymmetry; (c) {\it Mean Fragment Multiplicity} $Z \geq 3$.
The saturation of this curve defines the freeze-out configuration,
as we can also check from the density plots in Fig.2.
Bottom row, properties of the "primary" fragments in the
freeze-out configuration: (d) {\it Charge Distribution},
 (e) {\it Asymmetry Distribution} and
(f) {\it Fragment Multiplicity Distribution} (normalized to $1$).

We see a neutron dominated prompt particle emission and a second
{\it neutron burst} at the time of fragment formation in
the "central region". The latter is consistent with the dynamical
spinodal mechanism in dilute asymmetric nuclear matter, as
discussed before. The effect is quite reduced for semi-peripheral
collisions (compare Figs. 3b and 4b) and the IMF's produced in
the neck are more neutron rich (Figs. 3e and 4e).
This seems to indicate a different nature of the
fragmentation mechanism in central and neck regions, i.e.
a transition from volume to shape instabilities with different
isospin dynamics. In more peripheral collisions the interaction time
scale is also very reduced (Fig.4c) and this will
quench the isospin migration.

\section{Conclusions}

Starting from the thermodynamical features of instabilities and the 
dynamics of phase-transitions
in binary systems we obtain a consistent description of the 
multifragmentation process in heavy ion collisions and, in particular, 
of isospin effects. 
Isospin proves to be a useful probe in signaling a change in the fragment
formation mechanism passing from central to semipheripheral collisions.  
Moreover this "isospin dynamics" is found to be quite sensitive to the symmetry
term of the EOS, opening new stimulating perspectives on
such studies, which are of great astrophysical interest, under laboratory 
controlled conditions.

\section*{Acknowledgments}
One of authors (V.B.) gratefully acknowledges for the warm hospitality 
at LNS Catania during his Postdoctoral fellowship.

\end{document}